\documentclass[%
 reprint,
 amsmath,amssymb,
 aps,
]{revtex4-1}

\usepackage{graphicx}
\usepackage{dcolumn}
\usepackage{bm}
\usepackage{amsmath}

\begin{document}

\preprint{APS/123-QED}

\title{Nuclear-spin comagnetometer based on a liquid of identical molecules}

\author{Teng Wu,$^{1,*}$ John W. Blanchard,$^{1}$ Derek F. Jackson Kimball,$^{2}$ Min Jiang,$^{3}$ and Dmitry Budker$^{1,4}$}
\affiliation{    $^{1}$Helmholtz-Institut Mainz, Johannes Gutenberg University, 55128 Mainz, Germany\\
                    $^{2}$Department of Physics, California State University-East Bay, Hayward, California 94542-3084, USA\\
                    $^{3}$CAS Key Laboratory of Microscale Magnetic Resonance and Department of Modern Physics, University of Science and Technology of China, Hefei, Anhui 230026, China\\
                    $^{4}$Department of Physics, University of California at Berkeley, California 94720-7300, USA}
\date{\today}

\begin{abstract}
Atomic comagnetometers are used in searches for anomalous spin-dependent interactions.
Magnetic field gradients are one of the major sources of systematic errors in such experiments.
Here we describe a comagnetometer based on the nuclear spins within an ensemble of identical molecules.
The dependence of the measured spin-precession frequency ratio on the first-order magnetic field gradient is suppressed by over an order of magnitude compared to a comagnetometer based on overlapping ensembles of different molecules.
Our single-species comagnetometer is shown to be capable of measuring the hypothetical spin-dependent gravitational energy of nuclei at the $10^{-17}$ eV level, comparable to the most stringent existing constraints.
Combined with techniques for enhancing the signal such as parahydrogen-induced polarization, this method of comagnetometry offers the potential to improve constraints on spin-gravity coupling of nucleons by several orders of magnitude.
\end{abstract}

\maketitle

Atomic comagnetometers typically consist of overlapping ensembles of at least two different species of atomic spins \cite{Lamoreaux1986, Limes2018}.
The basic idea of comagnetometry is that the precession frequency of one spin can be used to monitor and compensate magnetic field fluctuations, while the other spin is used to search for nonmagnetic torques.
Practically, it is the ratio of the spin-precession frequencies of the different species under the influence of a bias magnetic field that is measured.
The ratio is relatively insensitive to changes in the magnetic field, but retains sensitivity to Zeeman-like nonmagnetic spin interactions.
Comagnetometers have been widely used for fundamental physics experiments \cite{Safronova2017}, such as measurements of permanent electric dipole moments (EDMs) \cite{Rosenberry2001, Regan2002, Baker2006, Griffith2009, Abel2017}, tests of $\it{CPT}$ and Lorentz invariance \cite{Bear2000, Altarev2009, Brown2010, Gemmel2010, Smiciklas2011, Smiciklas2011, Allmendinger2014}, and searches for exotic spin-dependent interactions mediated by hypothetical bosonic fields \cite{Venema1992, Vasilakis2009, Bulatowicz2013, Tullney2013, Hunter2013, Kimball2013, Heil2013, Kimball2017}.
Comagnetometers also find practical applications as sensitive gyroscopes \cite{Kornack2005, Jiang2018}.

In fundamental-physics experiments using comagnetometers based on overlapping ensembles of different species, one of the major systematic effects reducing accuracy is due to uncontrolled magnetic field gradients \cite{Baker2006, Kimball2017, Sheng2014}.
Previous work demonstrates that there exists some spatial separation between the ensemble-averaged position of different spin species due to nonuniform polarization \cite{Sheng2014}, gravity \cite{Baker2006}, and/or thermodiffusion effects \cite{Ledbetter2012}.
As a consequence, in the presence of a magnetic field gradient, the average magnetic field sensed by different spin species is different.
Thus the ratio of spin-precession frequencies acquires a magnetic-field-gradient dependence that can add noise and is difficult to distinguish from other sources of nonmagnetic torques on spins.
For this reason, complex arrangements are needed to monitor and reduce the magnetic field gradient for each cycle of measurement \cite{Limes2018, Kimball2017, Allmendinger2017}.

In contrast to comagnetometers which utilize overlapping ensembles of different atomic or molecular species, here we introduce and demonstrate a new comagnetometer configuration based on an ensemble of identical molecules.
In this single-species comagnetometer, different nuclear spins are probed within the same molecule.
In this way, the spatial sampling of the field by the different nuclear spins is made nearly identical and systematic errors related to field gradients are highly suppressed.
By taking advantage of the techniques of ultralow-field nuclear magnetic resonance (NMR) and sensitive atomic magnetometry, the $\it{J}$-coupling (indirect spin-spin coupling) spectrum of a liquid-state ensemble of acetonitrile-2-$^{13}$C molecules can be measured with sub-mHz precision in an ultralow magnetic field with a single scan (10 s measurement time).
Under the influence of a bias magnetic field, the $\it{J}$-coupling resonance lines at different frequencies split into separate peaks.
The frequency separation between the split peaks for each $\it{J}$-coupling resonance has distinct linear coefficients with respect to the magnetic field.
Measurements of these splittings can be employed as a comagnetometer.
We experimentally demonstrate that in the presence of a temperature gradient, such a comagnetometer is insensitive to first-order magnetic field gradients within experimental uncertainty.
We analyze a possible application of this new kind of comagnetometer for measurement of a coupling between nuclear spins and gravitational fields.

The device is based on a zero- to ultralow- field (ZULF) NMR configuration and the experimental setup is described in detail in Refs.~\cite{Ledbetter2012, Tayler2017}.
The spin ensemble we use to realize the comagnetometer is liquid-state acetonitrile-2-$^{13}$C ($^{13}$CH$_{3}$CN, from Sigma-Aldrich).
The acetonitrile-2-$^{13}$C sample ($\sim$~100 $\mu$L) is flame-sealed under vacuum in a standard 5 mm NMR tube.
The dissolved oxygen inside the sample is removed through a few freeze-pump-thaw cycles assisted with liquid nitrogen.
The sample is initially polarized in a 1.8 T Halbach magnet for about 30 s, and then pneumatically shuttled down into a four-layer magnetic shield (Twinleaf MS-1F).
During the transit of the sample, a $\sim$~30 $\mu$T magnetic field is applied to the sample with a solenoid, which is used to guide the initial spin magnetization along the vertical direction ($y$).
After the sample drops into the detection region ($\sim$~1 mm above a rubidium vapor cell), the guiding field is turned off within 10 $\mu$s.
The initial spin magnetization then evolves under the $\it{J}$-coupling interaction between $^{13}$C and the three $^{1}$H protons, which in turn generates an oscillating magnetization signal along $y$ and is detected with a rubidium atomic magnetometer (sensitivity $\approx$~10 fT/Hz$^{1/2}$).
A $\pi$ pulse ($\sim$~1 mT, 50 $\mu$s) for $^{13}$C along $x$ is triggered after switching off the guiding field and prior to data acquisition, which is tuned to provide the maximum signal amplitude.
Besides, a small bias field along $z$ is applied by coils within the innermost shield layer, and can be regarded as a small perturbation to the dominant $\it{J}$-coupling interaction.
The $\it{J}$-coupling spectrum is thus split into different peaks under the influence of the bias magnetic field.

\begin{figure}
\centering
\includegraphics[width=0.92\columnwidth]{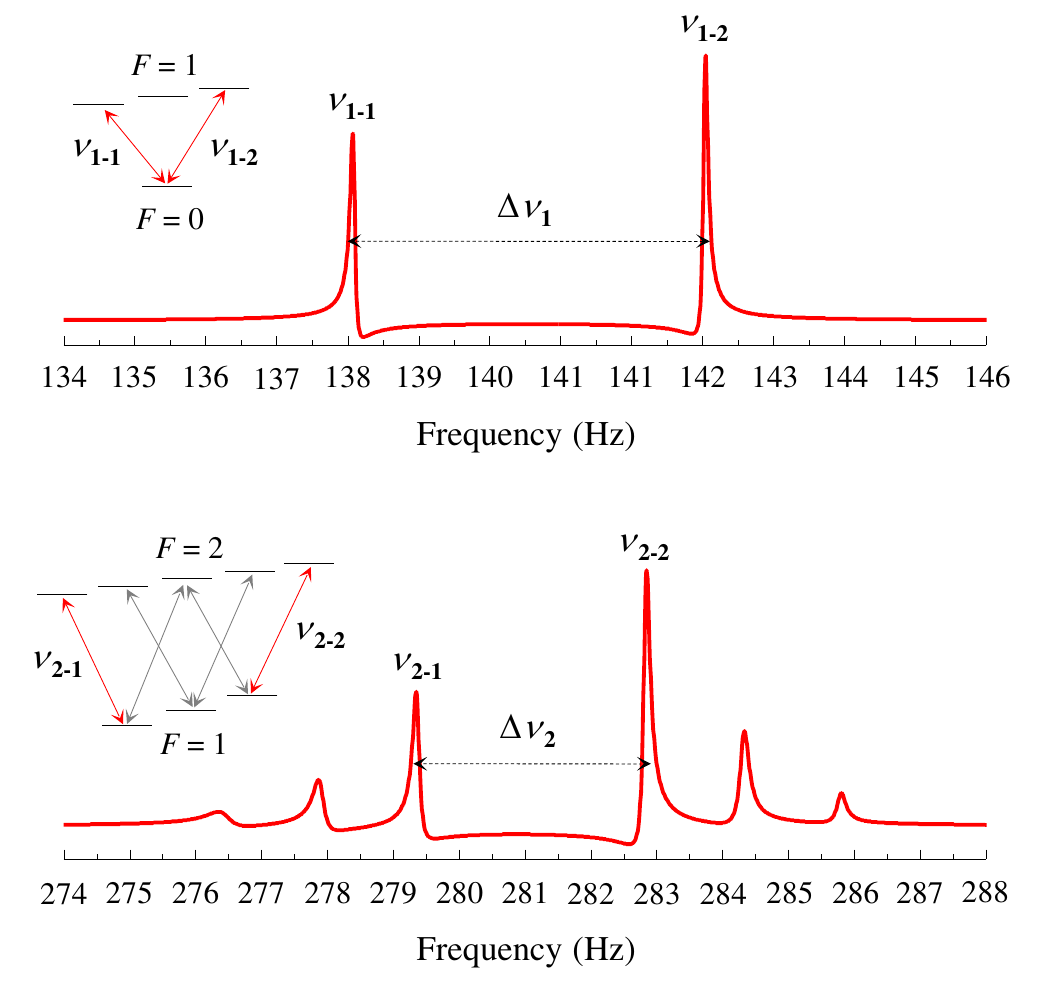}
\caption{(color online). Experimentally measured $\it{J}$-coupling spectrum of acetonitrile-2-$^{13}$C ($^{13}$CH$_{3}$CN) in a 80 nT bias field along $z$. The top and bottom traces show the split spectrum at $J_{\rm{CH}}$ and 2$J_{\rm{CH}}$, respectively. The related transitions used for comagnetometry are shown with solid red arrows.}
\end{figure}

\begin{figure}
\centering
\includegraphics[width=0.92\columnwidth]{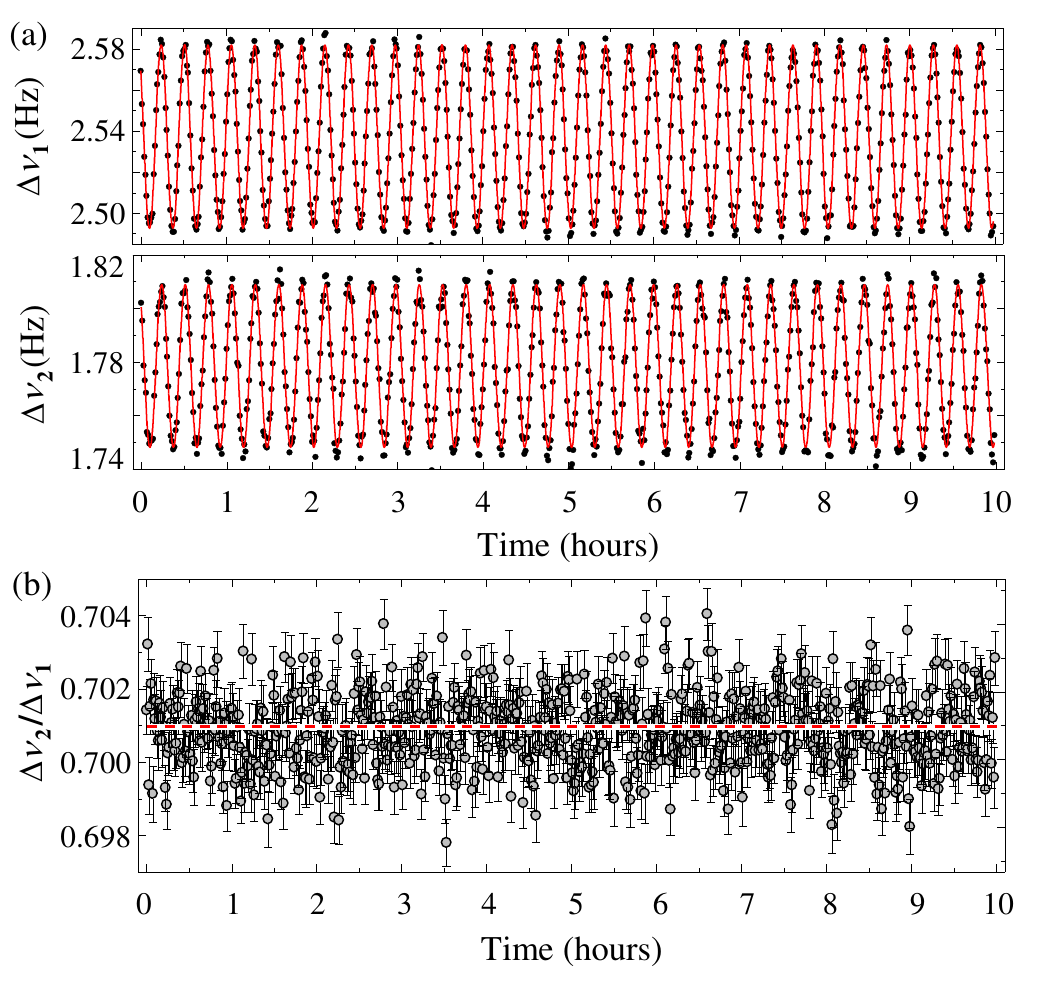}
\caption{(color online). Experimental demonstration of comagnetometry with an ensemble of identical molecules. (a) The measurements were taken while an oscillating magnetic field was applied in the $z$ direction (see text). The red solid lines are the fitted curves. (b) The calculated $\Delta\nu_{2}/\Delta\nu_{1}$ based on (a). The average value of $\Delta\nu_{2}/\Delta\nu_{1}$ (red dashed line) is 0.70088(4).}
\end{figure}

Acetonitrile-2-$^{13}$C is a $^{13}$CH$_{3}$ system with three equivalent protons.
Thus, the resulting zero-field $\it{J}$-coupling spectrum consists of two resonance lines, with one at $J_{\rm{CH}}$ and the other at 2$J_{\rm{CH}}$ \cite{Ledbetter2011, Blanchard2016}.
The measured $\it{J}$-coupling frequency for acetonitrile-2-$^{13}$C in our experiment is 140.55002(3) Hz, which is a function of the sample temperature (the frequency shift as a function of temperature is $\sim$~-125 $\mu$Hz/K).
With a small bias magnetic field ($\sim$~80 nT), the two lines split into different patterns of peaks, see Fig.~1.
The spectrum around $J_{\rm{CH}}$ splits into two peaks, while the spectrum around 2$J_{\rm{CH}}$ splits into six.
Within the 2$J_{\rm{CH}}$ multiplet, we focus on the central two peaks (the corresponding transitions are shown with solid red arrows in Fig.~1), as they have the highest signal-to-noise ratio compared to the others.
Neglecting all other nonmagnetic spin interactions, the frequencies for the two splittings $\Delta\nu_{1, 2}$ are $\Delta\nu_{1} = (\gamma_{h} + \gamma_{c})B_{z}$, $\Delta\nu_{2} = \frac{1}{2}(\gamma_{h} + 3\gamma_{c})B_{z}$, where $\gamma_{h, c}$ are the gyromagnetic ratios for $^{1}$H and $^{13}$C, respectively, and $B_{z}$ is the bias magnetic field \cite{Ledbetter2011, Appelt2010}.
There are no contributions from the second-order Zeeman effect on $\Delta\nu_{1, 2}$, and the third-order Zeeman effect is an order of magnitude smaller than the current experimental uncertainty (see Appendix B).
Since $\Delta\nu_{1}$ and $\Delta\nu_{2}$ are both proportional to $B_{z}$ but with different linear coefficients, we can employ them to realize a comagnetometer based on an ensemble of identical molecules.

\begin{figure}
\centering
\includegraphics[width=0.92\columnwidth]{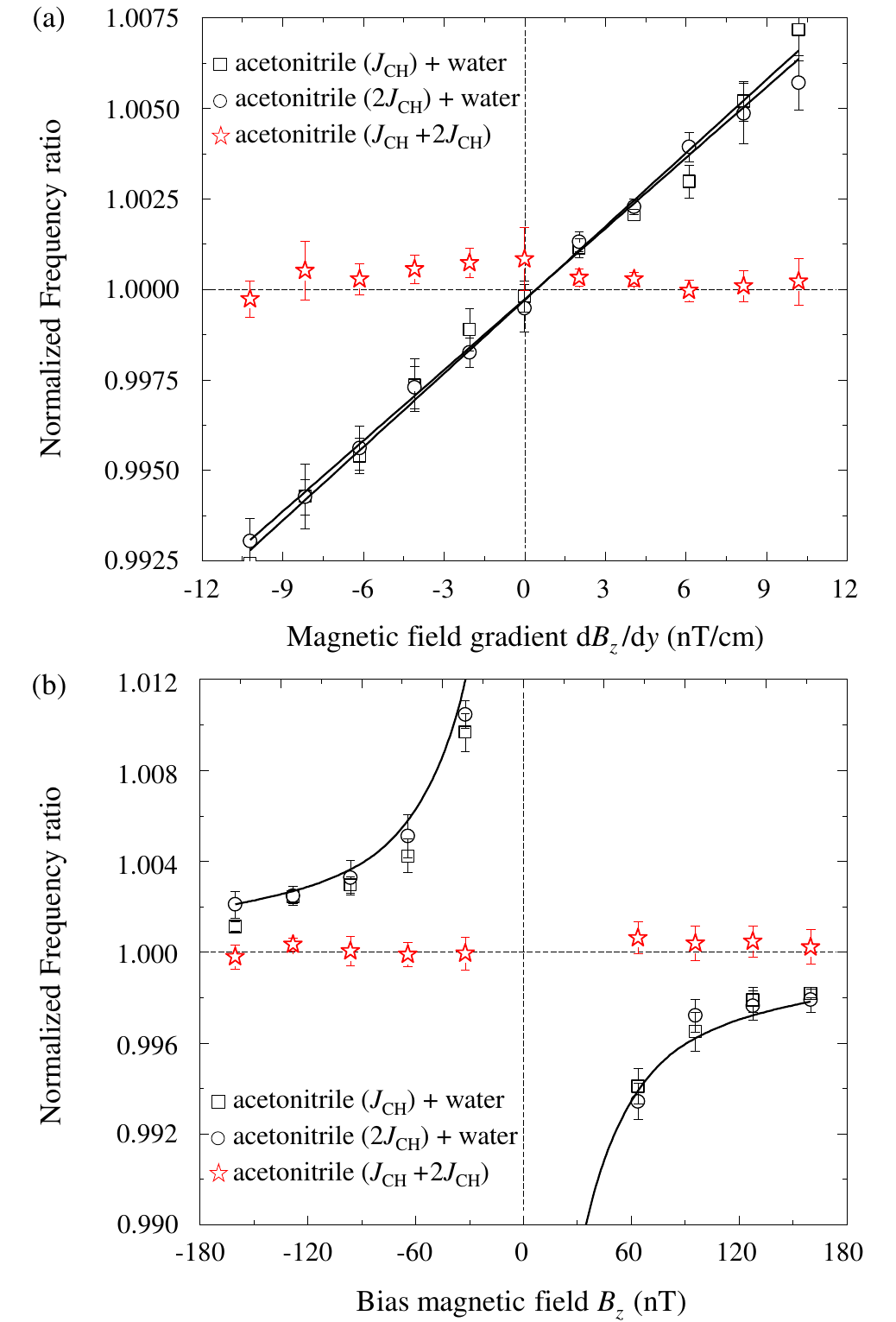}
\caption{(color online). Comparison of the normalized frequency ratios between a single-species comagnetometer (red star, $\Delta\nu_{2}/\Delta\nu_{1}$) and two dual-species reference comagnetometers (black square, $\nu_{h}/\Delta\nu_{1}$, black circle, $\nu_{h}/\Delta\nu_{2}$, discussed in the text). All the data are taken with the same acetonitrile-2-$^{13}$C (with $\sim$~1$\%$ water) and are normalized with the theoretical values at zero magnetic field gradient. (a) The normalized frequency ratio as a function of the gradient $\rm{d}\it{B}_{z}/\rm{d}\it{y}$, with constant bias magnetic field $B_{z} = 80$ nT. The solid lines are the linearly fitted curves. (b) The normalized frequency ratios as a function of bias magnetic fields $B_{z}$, with constant gradient $\rm{d}\it{B}_{z}/\rm{d}\it{y}$ = -3 nT/cm. The solid line is the fitted curve based on $B_{z}^{-1}$.}
\end{figure}

Comagnetometers should be able to suppress the variations in the bias magnetic field.
As a demonstration, we apply a slowly varying magnetic field along the same direction ($z$) as the bias field, with 1 mHz frequency and 0.5 nT amplitude.
Since the total acquisition time for each scan is 10 s, the oscillating magnetic field is effectively DC within this sampling window, and changes the two splitting frequencies by a few tens of mHz.
Figure~2(a) shows the measured frequencies $\Delta\nu_{1, 2}$ under the influence of the oscillating magnetic field, both of which display an evident 1 mHz modulation.
The ratio between $\Delta\nu_{1, 2}$ for each measurement is calculated and shown in Fig.~2(b).
Compared with Fig.~2(a), there is no apparent modulation of the frequency ratio.
Based on measurements over 10 hours, the averaged value of $\Delta\nu_{2}/\Delta\nu_{1}$ is 0.70088(4).
By using $\gamma_{h} = 42.5775$ MHz/T \cite {Mohr2008} and $\gamma_{c} = 10.7077$ MHz/T \cite{Antusek2005}, and taking into account the shielding factors of acetonitrile-2-$^{13}$C, i.e., $\sigma$($^{1}$H) = 31 ppm, $\sigma$($^{13}$C) = 185 ppm \cite{Antusek2005, Gottlieb1997}, the theoretical value is 0.70092.
Besides this, the third-order Zeeman effect modifies the frequency ratio at the level of $10^{-6}$ based on the current experimental parameters (see Appendix B).
However, systematic effects related to such a difference can be suppressed in precision measurements by employing field-reversal methods \cite{Kimball2017}.

\begin{figure*}
\centering
\includegraphics[width=1.7\columnwidth]{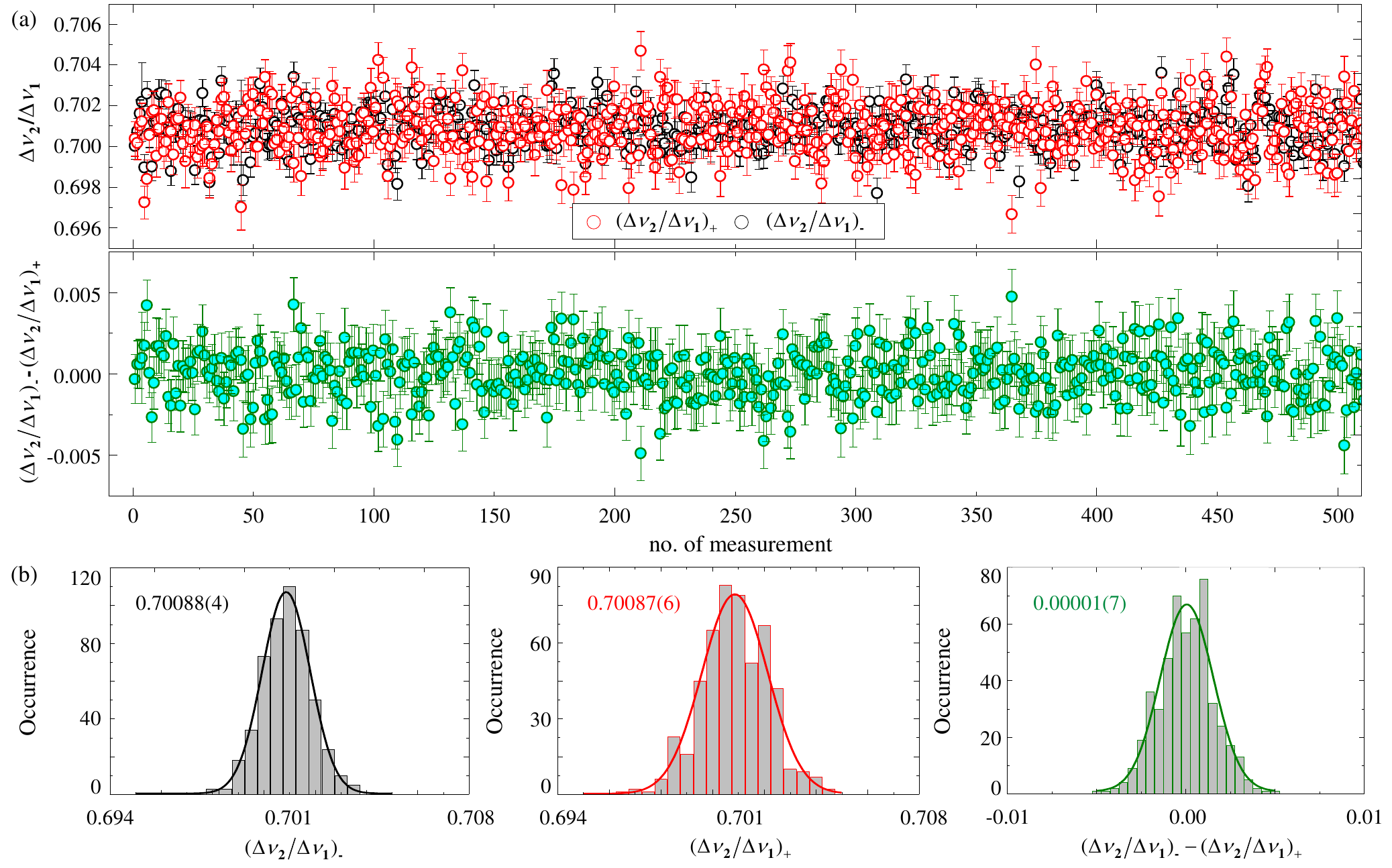}
\caption{(color online). Demonstration of the sensitivity and accuracy of a possible measurement of a spin-gravity coupling. (a) The measured frequency ratios $\mathcal{R}_{\pm} = \Delta\nu_{2}(\pm)/\Delta\nu_{1}(\pm)$ (top) and the calculated difference in $\Delta\nu_{2}/\Delta\nu_{1}$ based on the measured results of each consecutive $\{+B_{z}, -B_{z}\}$ (bottom). (b) Histograms of the results $(\Delta\nu_{2}/\Delta\nu_{1})_{-}$ (left), $(\Delta\nu_{2}/\Delta\nu_{1})_{+}$ (middle), and the difference (right). The numbers in the brackets are the standard error of the mean. The solid lines are the fitted curves using a Gaussian function, which indicate that the measurement results are consistent with normal distributions.}
\end{figure*}

It has been demonstrated, both experimentally and theoretically, that for a dual-species comagnetometer the spin-precession frequency ratio is a function of the magnetic field gradient.
A thorough investigation of magnetic-field-induced systematic effects can be found in Ref.~\cite{Sheng2014}.
Although their analysis is based on a gas-phase comagnetometer, many of the conclusions are also valid for a liquid-state comagnetometer.
Here, we focus on the shifts in the spin-precession frequency ratio due to the first-order gradient, typically, the gradient of the bias magnetic field along the vertical direction ($y$), i.e., $\rm{d}\it{B}_{z}/\rm{d}\it{y}$.
If there exist temperature gradients, different spin ensembles can experience different thermal diffusion rates, which in turn causes gradients in the concentration of the ensembles \cite{Sheng2014}.
Thus, a first-order magnetic field gradient can introduce an additional component in the frequency ratio, which has the form $G_{1}\Delta/B_{z}$, where $G_{1}$ is the first-order magnetic field gradient, $\Delta$ is the separation of the centers of the ensemble-averaged position of the spins.
Considering that our sample is placed at a small distance ($\sim$~1 mm) above the rubidium vapor cell, which is heated to $\sim170 ^{\circ}$C, there is a large temperature gradient along the vertical direction ($\sim$~25 K/cm).

In order to determine the sensitivity to magnetic field gradients, we compare our single-species comagnetometer to two dual-species reference comagnetometers.
The dual-species comagnetometers are based on the same acetonitrile-2-$^{13}$C, but use one of the splittings, $\Delta\nu_{1}$ or $\Delta\nu_{2}$, together with the precession frequency of $^{1}$H in residual water present in the sample ($\sim$~1$\%$).
The precession frequency of $^{1}$H can be written as $\nu_{h}=\gamma_{h}B_{z}$.
Therefore, for the two reference comagnetometers, the measured spin-precession frequency ratios are $\nu_{h}/\Delta\nu_{1}=\gamma_{h}/(\gamma_{h}+\gamma_{c})$ and $\nu_{h}/\Delta\nu_{2}=2\gamma_{h}/(\gamma_{h}+3\gamma_{c})$, respectively.

Figure 3(a) shows the spin-precession frequency ratios for the three comagnetometers as a function of the first-order magnetic field gradient $\rm{d}\it{B}_{z}/\rm{d}\it{y}$.
In order to compare the results at the same level, the measured spin-precession frequency ratios of each comagnetometer are normalized to the corresponding theoretical values at zero magnetic field gradient.
For the two reference comagnetometers, the normalized frequency ratios are both linear in the magnetic field gradient, with slopes of $6.71(22)\times 10^{-4}$ cm/nT ($\nu_{h}/\Delta\nu_{1}$, black squares) and $6.51(14) \times 10^{-4}$ cm/nT ($\nu_{h}/\Delta\nu_{2}$, black circles), respectively.
Such a linear dependence is also observed in Ref.~\cite{Ledbetter2012}, in which a mixture of pentane and hexafluorobenzene is used to realize a liquid-state comagnetometer.
The slopes of the normalized frequency ratios for the two reference comagnetometers are nearly identical since they are based on the same sample.
For the single-species comagnetometer, the results display a negligible linear dependence with the magnetic field gradient ($\Delta\nu_{2}/\Delta\nu_{1}$, red stars).
Fitting the results with a linear function gives a slope of $-0.24(24) \times 10^{-4}$ cm/nT, which is at least an order of magnitude smaller than the dual-species reference comagnetometers and, in fact, consistent with zero.
The residual nonlinear dependence could be attributed to higher order effects of the gradient on the precession frequencies, which could introduce broadening and shift of the resonance lines \cite{Sheng2014}.

Figure 3(b) shows the spin-precession frequency ratios as a function of the bias magnetic field.
We apply a constant gradient $\rm{d}\it{B}_{z}/\rm{d}\it{y}$ = -3 nT/cm through the gradient coils.
For the dual-species comagnetometer, the results are fit to the inverse of the bias magnetic field amplitude, $B_{z}^{-1}$, in agreement with the $G_{1}\Delta/B_{z}$ form of gradient dependence described above.
Under the same condition, there is no apparent dependence of the frequency ratio on $B_{z}^{-1}$ for the single-species comagnetometer.

We also apply first-order magnetic field gradients along $x$ and $z$ directions, i.e., $\rm{d}\it{B}_{z}/\rm{d}\it{x}$ and $\rm{d}\it{B}_{z}/\rm{d}\it{z}$.
Under these conditions, the frequency ratios measured with the reference comagnetometers similarly show no linear dependence on the first-order magnetic field gradient.
Since there are negligible temperature gradients along $x$ and $z$, the first-order gradient does not change the frequency ratio up to the second order of the gradient strength, if the Larmor frequency is much larger than the diffusion rate across the cell ($D/R^{2}$, $D$ is the diffusion constant, and $R$ is the cell radius) \cite{Cates1988, Sheng2014}.
This situation is well satisfied in our experiment, considering that the diffusion constants for acetonitrile and water are both on the order of $10^{-5} \rm{cm}^{2}/\rm{s}$, the radius of the tube is $\sim$ 0.2 cm, and the Larmor frequency is on the order of few Hz.
Therefore, the results presented in Fig.~3 confirm that, in the presence of a temperature gradient, the first-order magnetic field gradients can introduce systematic errors for a conventional dual-species comagnetometer, while they have a negligible effect on the single-species comagnetometer.

This new kind of single-species liquid-state comagnetometer can be applied to tests of fundamental physics.
One promising application is a search for a spin-gravity coupling.
Detailed discussions of possible spin-gravity couplings can be found in \cite{Safronova2017, Venema1992, Kimball2013, Heil2013, Kimball2017} and references therein.
Here we focus on the coupling of the nuclear spin to the gravitational field of the Earth.
A possible spin-gravity coupling to the $^{13}$C and $^{1}$H nuclei of acetonitrile-2-$^{13}$C can be parameterized as modifications of the spin-precession frequencies (see Appendix A)
\begin{align}
\Delta\nu_{1}(\pm) &= (\gamma_{h}+\gamma_{c})B_{z}\pm(-\frac{1}{3}\chi_{n}+\chi_{p})\frac{g\cos\phi}{\hbar}, \\
\Delta\nu_{2}(\pm) &= \frac{(\gamma_{h}+3\gamma_{c})B_{z}}{2}\pm(-\frac{1}{2}\chi_{n}+\frac{1}{2}\chi_{p})\frac{g\cos\phi}{\hbar}.
\end{align}
Here, $\pm$ refers to reversing the magnetic field direction, $\chi_{n}$ and $\chi_{p}$ are the gyrogravitational ratios of the neutron (from $^{13}$C) and proton (from $^{1}$H), respectively, $g$ is acceleration due to gravity, and $\phi$ is the angle between the magnetic field and the Earth's gravitational field \cite{Mayer1950, Klinkenberg1952, Kimball2015}.
We construct the ratio $\mathcal{R}_{\pm} \equiv \Delta\nu_{2}(\pm)/\Delta\nu_{1}(\pm)$.
The difference in the ratio obtained by reversing the magnetic field direction is
\begin{align}
\Delta\mathcal{R}\equiv\mathcal{R}_{-}\!-\!\mathcal{R}_{+}\thickapprox\frac{\gamma_{h}+3\gamma_{c}}{\gamma_{h}+\gamma_{c}}\left[\frac{(5\chi_{p}+4\chi_{n})g\cos\phi}{100\mu_{N}B_{z}}\right],
\end{align}
where $\mu_{N}$ is the nuclear magneton.

Due to the current system configuration, the angle $\phi$ is fixed at $90^{\circ}$ and can not be changed.
Thus, the contribution from the Earth's gravitational field is zero and can not be measured directly with our current system.
However, we can still reverse the magnetic field direction, and record the variations in $\Delta\mathcal{R}$, which demonstrates the achievable sensitivity for a measurement of the spin-gravity coupling.
Considering this, we reverse the magnetic field direction for each measurement scan, i.e., $+B_{z}, -B_{z}, +B_{z}, -B_{z},...$.
Each consecutive $\{+B_{z}, -B_{z}\}$ is taken as a group, for which $\Delta\mathcal{R}$ is calculated.
We perform 1024 continuous measurements ($\sim$~15 hours), which can be divided into 512 groups of $\{+B_{z}, -B_{z}\}$.
The measured frequency ratios $\mathcal{R}_{\pm}$ and the corresponding difference $\Delta\mathcal{R}$ are shown in Fig.~4(a), respectively, including the histograms for these results, see Fig.~4(b).
Based on these measurements, we find that $\Delta\mathcal{R} = (1 \pm 7_{\rm{stat}}) \times 10^{-5}$.
This uncertainty level indicates that for the current system, $(5\chi_{p}+4\chi_{n})$ could be measured at a level of $10^{-32}$ g cm, which probes the spin-dependent gravitational energy of a linear combination of the proton and neutron at a level of $10^{-17}$ eV.
This is comparable to the most stringent existing constraint on the spin-gravity coupling of protons \cite{Kimball2017}.

The measurement uncertainty for the current system is statistics-limited based on the signal-to-noise ratio of a single scan ($\sim$~100).
If instead of thermal polarization using a permanent magnet, hyperpolarization methods such as parahydrogen-induced polarization are employed \cite{Theis2011, Theis2012, Suefke2017}, it is possible to achieve more than a $10^{4}$ enhancement of the polarization.
This will enable a search for spin-gravity couplings of nuclei several orders of magnitude more sensitive than existing limits.
Moreover, we can also take advantage of high-sensitivity commercial atomic magnetometers, such as those from QuSpin Inc., which could make a new single-species comagnetometer more compact and easier to rotate.
Another advantage for this comagnetometer is that, by using different kinds of molecular samples, one can realize comagnetometers to search for spin-gravity couplings using various combinations of protons and neutrons.

In conclusion, we have demonstrated a new single-species liquid-state comagnetometer.
We have shown experimentally that the magnetic field gradient-induced systematic effects are significantly suppressed with a single-species comagnetometer as compared to a comagnetometer based on overlapping ensembles of different species.
We have introduced a proof-of-principle experiment for a spin-gravity coupling measurement.
Based on the current sensitivity, our system is already comparable to the most sensitive system for measuring the coupling of proton spins with Earth's gravitational field.
We have outlined the next steps for improving our comagnetometer based on parahydrogen-induced polarization and compact atomic magnetometers.
These improvements could facilitate the development of low-cost, high-precision, and robust table-top systems for long-term measurements of exotic spin-dependent interactions \cite{Safronova2017}.

This research was supported by the DFG Koselleck Program and the Heising-Simons and Simons Foundations, the European Research Council under the European Union's Horizon 2020 Research and Innovative Programme under Grant agreement No. 695405 (T. W., J. W. B., and D. B.), and by the National Science Foundation under Grant No. PHY-1707875 (D. F. J. K.). Correspondence and requests for materials should be addressed to T. W. (teng@uni-mainz.de).

\appendix

\section{Spin-gravity coupling}
This section presents detailed derivations of the equations (1)-(3) in the main text.

The Hamiltonian containing scalar, Zeeman and spin-gravity couplings can be written as \cite{Ledbetter2011, Kimball2013}
\begin{widetext}
\begin{align*}
H = H_{J}+H_{Z}+H_{G}=\hbar\sum_{j;k>j}J_{jk}\bold{I}_{j}\cdot\bold{I}_{k}+\hbar\sum_{i}\gamma_{i}\bold{I}_{i}\cdot\bold{B}+\sum_{i}\chi_{i}\bold{I}_{i}\cdot\bold{g}.\tag{S1}
\end{align*}
\end{widetext}
Here, $H_{J}$ is the scalar couplings of different nuclear spins, $H_{Z}$ is the Zeeman interaction, $H_{G}$	is the contribution from spin-gravity couplings,
$J_{jk}$ is the scalar coupling between spins $\bold{I}_{j}$ and $\bold{I}_{k}$,
$\gamma_{i}$ is the gyromagnetic ratio of spin $\bold{I}_{i}$,
$\bold{B}$ is the bias magnetic field,
$\chi_{i}$ is the gyrogravitational ratio for the $i$th spin,
and $\bold{g}$ is the gravitational acceleration due to Earth, which is the dominant gravitational field in a laboratory environment.
The spin-gravity coupling has the same form as the Zeeman interaction, and can be regarded as a quasi-magnetic field with a strength of
$\chi_{i}g\cos\phi/\hbar\gamma_{i}$, where $\phi$ is the angle between the bias magnetic field and the Earth's gravitational field.
If the bias magnetic field direction is reversed, this quasi-magnetic field changes its relative sign, and thus can be extracted by monitoring the variations of the Larmor precession frequencies.

Consider the case of acetonitrile-2-$^{13}$C, which is a $^{13}$CH$_{3}$ system with three equivalent protons.
In the absence of the Zeeman interaction and spin-gravity coupling, the unperturbed state $|f, m_{f}, k\rangle$ has energy \cite{Ledbetter2011}
\begin{align*}
E(f, k, s)=\hbar J/2[f(f+1)-k(k+1)-s(s+1)].\tag{S2}
\end{align*}
Here, $k=\sum_{i}k_{i}$ is the quantum number of the sum of the three equivalent proton spins, which has the values of 1/2 and 3/2, $s$ is the quantum number of the $^{13}$C spin and is 1/2,
$f$ is the quantum number of the total spin angular momentum, and has the values $f=0, 1$ for $k=1/2, s=1/2$, and $f=1,2$ for $k=3/2, s=1/2$.
Therefore, the zero-field $\it{J}$-coupling spectrum of acetonitrile-2-$^{13}$C consist of two resonance lines, with one at $E(1, 1/2, 1/2)-E(0,1/2,1/2)=J$ and the other one at $E(2, 3/2, 1/2)-E(1,3/2,1/2)=2J$.

In the limit where the Zeeman energies (and spin-gravity energy) are much smaller than the scalar couplings, we can use the first-order perturbation theory to calculate the shift in energy levels (higher-order effects are considered in the next section).
The eigenstates are still those of the unperturbed scalar Hamiltonian.
We can write the shifts in the energy levels due to Zeeman interaction and spin-gravity coupling as \cite{Ledbetter2011}
\begin{widetext}
\begin{align*}
\Delta E(f,m_{f},k,s) &= \langle fm_{f}|(H_{Z}+H_{G})|fm_{f}\rangle\\
&=\langle fm_{f}|[\hbar B_{z}(\gamma_{h}k_{z}+\gamma_{c}s_{z})+g\cos\phi(\chi_{h}k_{z}+\chi_{c}s_{z})]|fm_{f}\rangle\\
&=\sum_{m_{k},m_{s}}\langle ksm_{k}m_{s}|fm_{f}\rangle^{2}[\hbar B_{z}(\gamma_{h}m_{k}+\gamma_{c}m_{s})+g\cos\phi(\chi_{h}m_{k}+\chi_{c}m_{s})].\tag{S3}
\end{align*}
\end{widetext}
Here, $\langle ksm_{k}m_{s}|fm_{f}\rangle$ is the Clebsch-Gordan coefficient. $\gamma_{h}$ and $\gamma_{c}$ are the gyromagnetic ratios for $^{1}$H (proton) and $^{13}$C, respectively.
By using the selection rules of the magnetic dipole transitions \cite{Ledbetter2011}, i.e., $\Delta f=0, \pm 1, \Delta m_{f}=\pm 1, \Delta k=0$, under the influence of magnetic field, the spectrum around $\it{J}$ splits into two peaks, while the spectrum around 2$\it{J}$ splits into six, which are shown in Fig. 1 in the main text.

We now calculate the form of the two frequencies $\Delta\nu_{1, 2}$ that we employ to realize the comagnetometer. $\Delta\nu_{1}$ is the frequency difference of the transitions $|0,0, 1/2\rangle\to|1,1, 1/2\rangle$ and $|0,0, 1/2\rangle\to|1,-1, 1/2\rangle$, and $\Delta\nu_{2}$ is the frequency difference of the transitions $|1,1, 3/2\rangle\to|2,2, 3/2\rangle$ and $|1,-1, 3/2\rangle\to|2,-2, 3/2\rangle$.
Based on Eq.~(S3), the two frequencies $\Delta\nu_{1, 2}$ can be written as
\begin{align*}
&\Delta\nu_{1}(\pm)=(\gamma_{h}+\gamma_{c})B_{z}\pm(\chi_{c}+\chi_{h})\frac{g\cos\phi}{\hbar}, \\
&\Delta\nu_{2}(\pm)=\frac{(\gamma_{h}+3\gamma_{c})B_{z}}{2}\pm(\frac{3}{2}\chi_{c}+\frac{1}{2}\chi_{h})\frac{g\cos\phi}{\hbar}.\tag{S4}
\end{align*}
In Eqs.~(S4), $\pm$ refers to reversing the magnetic field direction.

The next step is to express the gyrogravitational ratios of $\chi_{c}$ and $\chi_{h}$ in terms of the coupling constants for the proton $\chi_{p}$ and neutron $\chi_{n}$.
Based on the nuclear shell model, for odd-A nuclei, the nuclear spin $\bold{I}$ is entirely due to the orbital motion and the intrinsic spin of the valence nucleon \cite{Mayer1950, Klinkenberg1952, Kimball2015}.
For $^{1}$H,  there is only a proton, whose state is $1s_{1/2}$. The $^{13}$C nucleus has a valence neutron, whose state is $2p_{1/2}$.
As do most theoretical models, we assume that there is no contribution from orbital angular momentum to the spin-gravity coupling \cite{Kimball2015}, and thus $\chi_{c}$ and $\chi_{h}$ can be rewritten as
\begin{align*}
&\chi_{h}=\frac{\langle\bold{S}_{p}\cdot\bold{I}_{h}\rangle}{I_{h}(I_{h}+1)}\chi_{p}=\chi_{p},\\
&\chi_{c}=\frac{\langle\bold{S}_{n}\cdot\bold{I}_{c}\rangle}{I_{c}(I_{c}+1)}\chi_{n}=-\frac{1}{3}\chi_{n}, \tag{S5}
\end{align*}
Here, $\bold{S}_{p}$ and $\bold{S}_{n}$ are the valence proton and neutron spins, $\bold{I}_{h}$ and $\bold{I}_{c}$ are the total angular momentum of $^{1}$H and $^{13}$C, respectively.
The quantum number of the orbital angular momentum for the valence proton of $^{1}$H is 0, and for the valence neutron of $^{13}$C, it is 1.
Replacing $\chi_{h}$ and $\chi_{c}$ in Eqs.~(S4), we obtain Eqs.~(1) and (2) in the main text.

Based on these, the spin-precession frequency ratio $\mathcal{R}\equiv\Delta\nu_{2}/\Delta\nu_{1}$ has the form
\begin{align*}
\mathcal{R}=\frac{\gamma_{h}+3\gamma_{c}}{2(\gamma_{h}+\gamma_{c})}\left[\frac{1+(-\chi_{n}+\chi_{p})\frac{g\cos\phi}{\hbar(\gamma_{h}+3\gamma_{c})B_{z}}}{1+(-\frac{1}{3}\chi_{n}+\chi_{p})\frac{g\cos\phi}{\hbar(\gamma_{h}+\gamma_{c})B_{z}}}\right]. \tag{S6}
\end{align*}
Considering the fact that $\gamma_{c,h} B_{z}\gg g\chi_{n,p}$, and replacing $\gamma_{h,c}$ with $g_{h,c}\mu_{N}/\hbar$, where $g_{h}\approx 5.6$ is the g-factor for $^{1}$H, and $g_{c}\approx 1.4$ is the g-factor for $^{13}$C, $\mu_{N}$ is the nuclear magneton, Eq.~(S6) can be rewritten as
\begin{align*}
\mathcal{R}\approx\frac{\gamma_{h}+3\gamma_{c}}{2(\gamma_{h}+\gamma_{c})}\left[1-\frac{(5\chi_{p}+4\chi_{n})g\cos\phi}{100\mu_{N}B_{z}}\right]. \tag{S7}
\end{align*}
Therefore, by reversing the magnetic field, $\Delta\mathcal{R}\equiv\mathcal{R}_{+}-\mathcal{R}_{-}$ can be calculated, which has the form as Eq.~(3) shown in the main text.

\section{Higher-order Zeeman effects}
This section shows the energy shifts of the magnetic sublevels due to higher-order Zeeman effects.

Based on the previous work \cite{Appelt2010}, for the $^{13}$CH$_{3}$ system with three equivalent protons, under the influence of a magnetic field, the energy of a sublevel can be expressed as
\begin{widetext}
\begin{align*}
E(f,m_{f},k,s)/\hbar=m_{f}\gamma_{h}B_{z}-J/4+s\sqrt{(k+1/2)^{2}J^{2}+2m_{f}J(\gamma_{c}-\gamma_{h})B_{z}+(\gamma_{c}-\gamma_{h})^{2}B_{z}^2}, \tag{S8}
\end{align*}
\end{widetext}
where all the symbols have the same definitions as in Eqs. (S2) and (S3).
Equation (S8) is valid for arbitrary fields $B_{z}$, and is the same as Eq. (S2) when $B_{z}=0$.

As mentioned before, $\Delta\nu_{1}$ is the frequency difference of the transitions $|0,0, 1/2\rangle\to|1,1, 1/2\rangle$ and $|0,0, 1/2\rangle\to|1,-1, 1/2\rangle$.
Based on Eq. (S8), the energies of these states can be calculated as
\begin{align*}
&E(0,0,1/2,1/2)/\hbar=-\frac{J}{4}-\frac{1}{2}\sqrt{J^{2}+(\gamma_{c}-\gamma_{h})^{2}B_{z}^2},\\
&E(1,1,1/2,1/2)/\hbar=\frac{J}{4}+\frac{1}{2}(\gamma_{c}+\gamma_{h})B_{z},\\
&E(1,-1,1/2,1/2)/\hbar=\frac{J}{4}-\frac{1}{2}(\gamma_{c}+\gamma_{h})B_{z},\tag{S9}
\end{align*}
Since $|1,1, 1/2\rangle$ and $|1,-1, 1/2\rangle$ are streched states, there are no higher-order Zeeman effects for these states.
Based on Eq. (S9), and the definition of $\Delta\nu_{1}$, we can have $\Delta\nu_{1}=(\gamma_{c}+\gamma_{h})B_{z}$,
which is linear with the magnetic field $B_{z}$.

We can also calculate $\Delta\nu_{2}$, which is the frequency difference of the transitions $|1,1, 3/2\rangle\to|2,2, 3/2\rangle$ and $|1,-1, 3/2\rangle\to|2,-2, 3/2\rangle$.
Similarly, the energy levels for these states are
\begin{widetext}
\begin{align*}
&E(1,-1,3/2,1/2)/\hbar=-\gamma_{h}B_{z}-\frac{J}{4}-\frac{1}{2}\sqrt{4J^{2}-2J(\gamma_{c}-\gamma_{h})B_{z}+(\gamma_{c}-\gamma_{h})^{2}B_{z}^2},\\
&E(2,-2,3/2,1/2)/\hbar=\frac{3J}{4}-\frac{1}{2}(\gamma_{c}+3\gamma_{h})B_{z},\\
&E(1,1,3/2,1/2)/\hbar=\gamma_{h}B_{z}-\frac{J}{4}-\frac{1}{2}\sqrt{4J^{2}+2J(\gamma_{c}-\gamma_{h})B_{z}+(\gamma_{c}-\gamma_{h})^{2}B_{z}^2},\\
&E(2,2,3/2,1/2)/\hbar=\frac{3J}{4}+\frac{1}{2}(\gamma_{c}+3\gamma_{h})B_{z}.\tag{S10}
\end{align*}
\end{widetext}
Since $|2,2, 3/2\rangle$ and $|2,-2, 3/2\rangle$ are streched states, there are no higher-order Zeeman effects for these states.
By performing a Taylor expansion for the square root in Eq. (S10) up to the third order in the small parameter $x=(\gamma_{c}-\gamma_{h})B_{z}/2J$, the energy levels for states $|1,-1, 3/2\rangle$ and $|1,1, 3/2\rangle$ can be written as
\begin{widetext}
\begin{align*}
&E(1,-1,3/2,1/2)/\hbar=-\frac{5J}{4}-\frac{(5\gamma_{h}-\gamma_{c})B_{z}}{4}-\frac{3(\gamma_{c}-\gamma_{h})^{2}B_{z}^2}{32J}-\frac{3(\gamma_{c}-\gamma_{h})^{3}B_{z}^{3}}{128J^{2}},\\
&E(1,1,3/2,1/2)/\hbar=-\frac{5J}{4}+\frac{(5\gamma_{h}-\gamma_{c})B_{z}}{4}-\frac{3(\gamma_{c}-\gamma_{h})^{2}B_{z}^2}{32J}+\frac{3(\gamma_{c}-\gamma_{h})^{3}B_{z}^{3}}{128J^{2}}.\tag{S11}
\end{align*}
\end{widetext}
Based on these, $\Delta\nu_{2}$ is calculated as $B_{z}(3\gamma_{c}+\gamma_{h})/2+3(\gamma_{c}-\gamma_{h})^{3}B_{z}^{3}/64J^{2}$.
We find that the even-order Zeeman effects shift the energy levels along the same direction, which in turn have no contributions in $\Delta\nu_{2}$.
Therefore, the frequency ratio $\mathcal{R}\equiv\Delta\nu_{2}/\Delta\nu_{1}$ can be written as
\begin{align*}
\mathcal{R}=\frac{\gamma_{h}+3\gamma_{c}}{2(\gamma_{h}+\gamma_{c})}+\frac{3(\gamma_{c}-\gamma_{h})^{3}B_{z}^{2}}{64J^{2}(\gamma_{h}+\gamma_{c})}.\tag{S12}
\end{align*}
Using the parameters in our experiment, i.e., $B_{z}=80$ nT, and $J=140$ Hz, the second term in Eq. (S12) is $8\times 10^{-6}$, which is smaller than the current experimental uncertainty.
For the spin-gravity coupling measurement, the systematic effects in $\Delta\mathcal{R}\equiv\mathcal{R}_{-}-\mathcal{R}_{+}$ due to the third-order Zeeman effect could be suppressed through reducing the magnetic field, employing field-reversal methods, and using NMR samples with higher $\it{J}$-coupling frequencies.


\begin{thebibliography}{0}%
\makeatletter
\providecommand \@ifxundefined [1]{%
 \@ifx{#1\undefined}
}%
\providecommand \@ifnum [1]{%
 \ifnum #1\expandafter \@firstoftwo
 \else \expandafter \@secondoftwo
 \fi
}%
\providecommand \@ifx [1]{%
 \ifx #1\expandafter \@firstoftwo
 \else \expandafter \@secondoftwo
 \fi
}%
\providecommand \natexlab [1]{#1}%
\providecommand \enquote  [1]{``#1''}%
\providecommand \bibnamefont  [1]{#1}%
\providecommand \bibfnamefont [1]{#1}%
\providecommand \citenamefont [1]{#1}%
\providecommand \href@noop [0]{\@secondoftwo}%
\providecommand \href [0]{\begingroup \@sanitize@url \@href}%
\providecommand \@href[1]{\@@startlink{#1}\@@href}%
\providecommand \@@href[1]{\endgroup#1\@@endlink}%
\providecommand \@sanitize@url [0]{\catcode `\\12\catcode `\$12\catcode
  `\&12\catcode `\#12\catcode `\^12\catcode `\_12\catcode `\%12\relax}%
\providecommand \@@startlink[1]{}%
\providecommand \@@endlink[0]{}%
\providecommand \url  [0]{\begingroup\@sanitize@url \@url }%
\providecommand \@url [1]{\endgroup\@href {#1}{\urlprefix }}%
\providecommand \urlprefix  [0]{URL }%
\providecommand \Eprint [0]{\href }%
\providecommand \doibase [0]{http://dx.doi.org/}%
\providecommand \selectlanguage [0]{\@gobble}%
\providecommand \bibinfo  [0]{\@secondoftwo}%
\providecommand \bibfield  [0]{\@secondoftwo}%
\providecommand \translation [1]{[#1]}%
\providecommand \BibitemOpen [0]{}%
\providecommand \bibitemStop [0]{}%
\providecommand \bibitemNoStop [0]{.\EOS\space}%
\providecommand \EOS [0]{\spacefactor3000\relax}%
\providecommand \BibitemShut  [1]{\csname bibitem#1\endcsname}%
\let\auto@bib@innerbib\@empty
\end{thebibliography}%


\begin{thebibliography}{99}

\bibitem{Lamoreaux1986} S. K. Lamoreaux, J. P. Jacobs, B. R. Heckel, F. J. Raab, and E. N. Fortson, Phys. Rev. Lett. $\bf{57}$, 3125 (1986).
\bibitem{Limes2018} M. E. Limes, D. Sheng, and M. V. Romalis, Phys. Rev. Lett. $\bf{120}$, 033401 (2018).
\bibitem{Safronova2017} M. S. Safronova, D. Budker, D. DeMille, D. F. Jackson Kimball, A. Derevianko, and C. W. Clark, arxiv: 1710.01833 (2017).
\bibitem{Rosenberry2001} M. A. Rosenberry and T. E. Chupp, Phys. Rev. Lett. $\bf{86}$, 22 (2001).
\bibitem{Regan2002} B. C. Regan, E. D. Commins, C. J. Schmidt, and D. DeMille, Phys. Rev. Lett. $\bf{88}$, 071805 (2002).
\bibitem{Baker2006} C. A. Baker, D. D. Doyle, P. Geltenbort, K. Green, M. G. D. van der Grinten, P. G. Harris, P. Iaydjiev, S. N. Ivanov, D. J. R. May, J. M. Pendlebury, J. D. Richardson, D. Shiers, and K. F. Smith, Phys. Rev. Lett. $\bf{97}$, 131801 (2006).
\bibitem{Griffith2009} W. C. Griffith, M. D. Swallows, T. H. Loftus, M.V. Romalis, B. R. Heckel, and E. N. Fortson, Phys. Rev. Lett. $\bf{102}$, 101601 (2009).
\bibitem{Abel2017} C. Abel $\it{et~al}.$, Phys. Rev. X $\bf{7}$, 041034 (2017).
\bibitem{Bear2000} D. Bear, R. E. Stoner, R. L. Walsworth, V. A. Kosteleck$\acute{\rm{y}}$, and C. D. Lane, Phys. Rev. Lett. $\bf{85}$, 5038 (2000).
\bibitem{Altarev2009} I. Altarev, W. Tian, A. Kreyssig, J. L. Zarestky, S. Nandi, N. Ni, S. L. Bud’ko, P. C. Canfield, A. I. Goldman, and R. J. McQueeney, Phys. Rev. Lett. $\bf{103}$, 081602 (2009).
\bibitem{Brown2010} J. M. Brown, S. J. Smullin, T.W. Kornack, and M. V. Romalis, Phys. Rev. Lett. $\bf{105}$, 151604 (2010).
\bibitem{Gemmel2010} C. Gemmel, W. Heil, S. Karpuk, K. Lenz, Ch. Ludwig, Yu. Sobolev, K. Tullney, M. Burghoff, W. Kilian, S. Knappe-Gr$\ddot{\rm{u}}$neberg, W. M$\ddot{\rm{u}}$ller, A. Schnabel, F. Seifert, L. Trahms, and St. Bae$\ss$ler, Eur. Phys. J. D $\bf{57}$, 303 (2010).
\bibitem{Smiciklas2011} M. Smiciklas, J. M. Brown, L.W. Cheuk, S. J. Smullin, and M. V. Romalis, Phys. Rev. Lett. $\bf{107}$, 171604 (2011).
\bibitem{Allmendinger2014} F. Allmendinger, W. Heil, S. Karpuk, W. Kilian, A. Scharth, U. Schmidt, A. Schnabel, Yu. Sobolev, and K. Tullney, Phys. Rev. Lett. $\bf{112}$, 110801 (2014).
\bibitem{Venema1992} B. J. Venema, P. K. Majumder, S. K. Lamoreaux, B. R. Heckel, and E. N. Fortson, Phys. Rev. Lett. $\bf{68}$, 135 (1992).
\bibitem{Vasilakis2009} G. Vasilakis, J. M. Brown, T. W. Kornack, and M. V. Romalis, Phys. Rev. Lett. $\bf{103}$, 261801 (2009).
\bibitem{Bulatowicz2013} M. Bulatowicz, R. Griffith, M. Larsen, J. Mirijanian, C. B. Fu, E. Smith, W. M. Snow, H. Yan, and T. G. Walker, Phys. Rev. Lett. $\bf{111}$, 102001 (2013).
\bibitem{Tullney2013} K. Tullney, F. Allmendinger, M. Burghoff, W. Heil, S. Karpuk, W. Kilian, S. Knappe-Gr$\ddot{\rm{u}}$neberg, W. M$\ddot{\rm{u}}$ller, U. Schmidt, A. Schnabel, F. Seifert, Yu. Sobolev, and L. Trahms, Phys. Rev. Lett. $\bf{111}$, 100801 (2013).
\bibitem{Hunter2013} L. Hunter, J. Gordon, S. Peck, D. Ang, and J. Lin, Science $\bf{339}$, 928 (2013).
\bibitem{Kimball2013} D. F. Jackson Kimball, I. Lacey, J. Valdez, J. Swiatlowski. C. Rios, R. Peregrina‐Ramirez, C. Montcrieffe, J. Kremer, J. Dudley, and C. Sanchez, Ann. Phys. (Berlin) $\bf{525}$, 514 (2013).
\bibitem{Heil2013} W. Heil, C. Gemmel, S. Karpuk, Y. Sobolev, K. Tullney, F. Allmendinger, U. Schmidt, M. Burghoff, W. Kilian, S. Knappe-Gr$\ddot{\rm{u}}$neberg, A. Schnabel, F. Seifert, and L. Trahms, Ann. Phys. (Berlin) $\bf{525}$, 539 (2013).
\bibitem{Kimball2017} D. F. Jackson Kimball, J. Dudley, Y. Li, D. Patel, and J. Valdez, Phys. Rev. D $\bf{96}$, 075004 (2017).
\bibitem{Kornack2005} T. W. Kornack, R. K. Ghosh, and M.V. Romalis, Phys. Rev. Lett. $\bf{95}$, 230801 (2005).
\bibitem{Jiang2018} L. Jiang, W. Quan, R. Li, W. Fan, F. Liu, J. Qin, S. Wan, and J. Fang, App. Phys. Lett. $\bf{112}$, 054103 (2018).
\bibitem{Sheng2014} D. Sheng, A. Kabcenell, and M. V. Romalis, Phys. Rev. Lett. $\bf{113}$, 163002 (2014).
\bibitem{Ledbetter2012} M. P. Ledbetter, S. Pustelny, D. Budker, M. V. Romalis, J. W. Blanchard, and A. Pines, Phys. Rev. Lett. $\bf{108}$, 243001 (2012).
\bibitem{Allmendinger2017} F. Allmendinger, P. Bl$\ddot{\rm{u}}$mler, M. Doll, O. Grasdijk, W. Heil, K. Jungmann, S. Karpuk, H.-J. Krause, A. Offenh$\ddot{\rm{a}}$usser, M. Repetto, U. Schmidt, Y. Sobolev, K. Tullney, L. Willmann, and S. Zimmer, Eur. Phys. J. D $\bf{71}$, 98 (2017).
\bibitem{Tayler2017} M. C. D. Tayler, T. Theis,  T. F. Sjolander, J. W. Blanchard, A. Kentner, S. Pustelny, A. Pines, and D. Budker, Rev. Sci. Instrum. $\bf{88}$, 091101 (2017).
\bibitem{Ledbetter2011} M. P. Ledbetter, T. Theis, J.W. Blanchard, H. Ring, P. Ganssle, S. Appelt, B. Bl$\ddot{\rm{u}}$mich, A. Pines, and D. Budker, Phys. Rev. Lett. $\bf{107}$, 107601 (2011).
\bibitem{Blanchard2016} J. W. Blanchard, D. Budker, eMagRes $\bf{5}$, 1395 (2016).
\bibitem{Appelt2010} S. Appelt, F. W. H$\ddot{\rm{a}}$sing, U. Sieling, A. Gordji-Nejad, S. Gl$\ddot{\rm{o}}$ggler, and B. Bl$\ddot{\rm{u}}$mich, Phys. Rev. A $\bf{81}$, 023420 (2010).
\bibitem{Mohr2008} P. J. Mohr, B. N. Taylor, and D. B. Newell, Rev. Mod. Phys. $\bf{80}$, 633 (2008).
\bibitem{Antusek2005} A. Antu$\breve{\rm{s}}$ek, K. Jackowski, M. Jaszu$\acute{\rm{n}}$ski, W. Makulski, and M. Wilczek, Chem. Phys. Lett. $\bf{411}$, 111 (2005).
\bibitem{Gottlieb1997} H. E. Gottlieb, V. Kotlyar, and A. Nudelman, J. Org. Chem. $\bf{62}$, 7512 (1997).
\bibitem{Cates1988} G. D. Cates, S. R. Schaefer, and W. Happer, Phys. Rev. A $\bf{37}$, 2877 (1988).
\bibitem{Mayer1950} M. G. Mayer, Phys. Rev. $\bf{78}$, 16 (1950).
\bibitem{Klinkenberg1952} P. F. A. Klinkenberg, Rev. Mod. Phys. $\bf{24}$, 63 (1952).
\bibitem{Kimball2015} D. F. Jackson Kimball, New J. Phys. $\bf{17}$, 073008 (2015).
\bibitem{Theis2011} T. Theis, P. Ganssle, G. Kervern, S. Knappe, J. Kitching, M. P. Ledbetter, D. Budker, and A. Pines, Nat. Phys. $\bf{7}$, 571 (2011).
\bibitem{Theis2012} T. Theis, M. P. Ledbetter, G. Kervern, J. W. Blanchard, P. J. Ganssle, M. C. Butler, H. D. Shin, D. Budker, and A. Pines, J. Am. Chem. Soc. $\bf{134}$, 3987 (2012).
\bibitem{Suefke2017}M. Suefke, S. Lehmkuhl, A. Liebisch, B. Bl$\ddot{\rm{u}}$mich, and S. Appelt, Nat. Phys. $\bf{13}$, 568 (2017).
\end{thebibliography}
\end{document}